\documentclass[letter]{aa}

\usepackage{color} 

\usepackage{ulem} 

\usepackage{graphicx}
\usepackage{amsmath,epsfig,rotating,amssymb}

\begin{document}

\def \cha{\widehat}
\def \div{{\rm div}}
\def \grad{{\rm {\bf grad} }}
\def \kms{{\rm ~km~s}^{-1}}
\def \pcc{{\rm ~cm}^{-3}}
\def \psc{{\rm ~cm}^{-2}}
\def \pr{{\it permanent}  regime }
\def \rot{{\rm {\bf rot} }}
\def \VS{{V\'azquez-Semadeni}}


\author{Hennebelle P. \inst{1}, Banerjee R. \inst{2},
V\'azquez-Semadeni E. \inst{3}, Klessen R. \inst{2}, Audit E. \inst{4}}

\institute{Laboratoire de radioastronomie, UMR 8112 du
CNRS, 
\newline {\'E}cole normale sup{\'e}rieure et Observatoire de Paris,
 24 rue Lhomond, 75231 Paris cedex 05, France 
\and Zentrum fuer Astronomie der Universitat Heidelberg, Institut 
fuer Theoretische Astrophysik, 69120 Heidelberg, Germany 
 \and Centro de Radioastronom\' \i a y Astrof\'\i sica, Universidad Aut\'onoma de M\'exico, Apdo Postal 3-72 Morelia, 58089, M\'exico
\and  Service d'Astrophysique, CEA/DSM/DAPNIA/SAp, Centre d'\'Etudes de Saclay,
l'Orme les Merisiers, 91191 Gif-sur Yvette Cedex, France}

\offprints{  P. Hennebelle \\
{\it patrick.hennebelle@ens.fr}   }

\title{From the warm magnetized atomic medium to molecular clouds}

\authorrunning{Hennebelle et al.}


\abstract
{ It has  recently been proposed that giant molecular
complexes form at the sites where streams of diffuse warm
atomic gas collide at transonic velocities.}
{We study the  global statistics of molecular clouds formed by large scale
   colliding    flows
of warm neutral atomic interstellar  gas under
ideal MHD conditions.  The flows deliver material as well
as  kinetic energy and trigger thermal instability leading eventually to 
gravitational collapse.}
{We perform adaptive  mesh refinement MHD simulations which,  for
      the first time in this context, treat self-consistently cooling and self-gravity.}
{The clouds formed in the simulations  develop a highly
inhomogeneous density and temperature structure, with cold dense
filaments and clumps
condensing from converging flows of warm atomic gas. In the clouds, the
column density probability density distribution (PDF) peaks at $\sim 2
\times 10^{21} \psc$ and decays rapidly at higher
values; the magnetic intensity correlates weakly with density from $n
\sim 0.1$ to $10^4 \pcc$, and then varies roughly as $n^{1/2}$ for higher
densities. }
{The global statistical properties of such molecular
    clouds are reasonably consistent with
observational determinations.
 Our numerical simulations suggest that molecular clouds
formed by the moderately supersonic collision of warm atomic gas streams.}

\maketitle

\section{Introduction}
The formation of molecular clouds is one of the key steps for the star
formation process.  A large number of studies investigate the
internal dynamics of molecular clouds (e.g. see the review by MacLow
\& Klessen 2004), but only a few investigations address the problem of
their formation itself  (e.g. see the review by
Hennebelle, Mac Low \& V\'azquez-Semadeni  2008). This is partly due to the difficulty in
treating the large range of spatial scales relevant in this problem
and partly due to uncertainties on the mechanisms at the origin of
their formation.  During the last decade, the idea has emerged that
the molecular clouds may be formed at the onset of a large scale
converging flow of atomic gas (e.g. Ballesteros-Paredes et al. 1999)
with the active role of thermal instability (Hennebelle \& P\'erault
1999, Koyama \& Inutsuka 2000, 2002, Audit \& Hennebelle 2005, Heitsch
et al. 2005, V\'azquez-Semadeni et al. 2006).  Indeed, since the
density of molecular clouds is much larger than the mean ISM density,
 large scale flows are a viable explanation to excite density
enhancements.
The origin  
of these flows is however unclear and  may be not unique.
 Most likely they  arise from turbulent fluctuations 
or gravitational instability occuring at  large scales. 

Recently large multi-dimensional   non-magnetic
 numerical simulations   have been
performed (Hennebelle \& Audit 2007, V\'azquez-Semadeni et al. 2007, 
Heitsch et al. 2008, the last two including self-gravity)
 with the aim of studying in  detail the formation of 
dense gas from a large flow of warm neutral medium,
resolving down to the star forming scales.

In this letter, we present 
the first results of large adaptive mesh refinement (AMR), MHD simulations
performed with the codes RAMSES (Teyssier 2002, Fromang et al. 2006) 
and FLASH (Fryxell et al. 2000). 
These   are the first simulations which, starting from the WNM, 
include  magnetic
field, cooling, self-gravity, and thanks to the AMR scheme,  have 
sufficient spatial resolution  to resolve individual, high
  density, clouds.  
The molecular clouds we observe in these simulations are
  self-consistently generated by thermal instabilities, turbulence 
and  gravitational contraction. 
We stress that  
the internal structure and the
  turbulent properties of these molecular clouds 
are not the result of  an ad hoc assumption on the external
  turbulent driving.  By performing these 
numerical simulations, we expect to tackle  unsolved  and outstanding 
questions such  as what drives turbulence in molecular
clouds; what is the gas density and temperature distribution, 
and what is the structure of the magnetic field in these objects{\bf ?}
In this letter we  report on the most important global
properties of the clouds. In subsequent articles (Banerjee et al. 2008),  
we will give a more
detailed analysis of the cloud formation, structure and evolution,
and on the efficiency of star formation in the simulations.

In section 2, we describe the numerical set up and the initial conditions
whereas in section 3 we present our results and  preliminary 
comparisons with observations. Section 4 concludes  this paper.

\begin{figure}
\includegraphics[width=6.5cm]{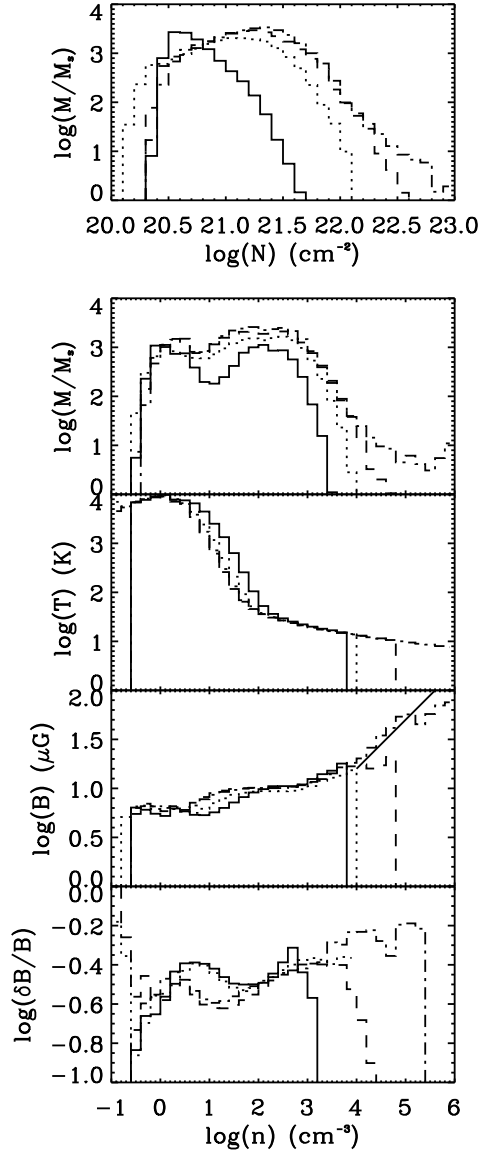}
\caption{Top and second panels: Column density and density PDF 
in the simulation. 
Third, fourth and fifth panels: Temperature, 
magnetic intensity and magnetic intensity variance as a function of 
gas density.  The {\it solid, dotted, dashed, and
dash-dotted} lines show the distributions at times $t=3.7$, 7.7,
12.0 and 13.8 Myr, respectively. The straight line  in
the fourth panel 
shows a distribution proportional to  $n^{1/2}$. }
\label{histo}
\end{figure}

\section{Numerical setup and initial conditions}
The numerical  simulation presented in this letter  has
been performed with  
 the AMR code RAMSES using the HLL solver. Ramses is a second order
Godunov scheme and uses the constraint transport method to ensure
  ${\rm div} B =0$ (see Fromang et al. 2006).
Starting with an initial resolution of 256$^3$ cells, 2 levels of 
refinement are allowed during the calculation leading to an effective 
1024$^3$ numerical simulation. The criterion used to refine the grid is a 
simple density threshold of 50 cm$^{-3}$ for the first level and 200 cm$^{-3}$
for the second one. This ensures that the dense gas is uniformly 
resolved.  With the box size being about 50 pc, this leads to 
 a spatial resolution of about 0.05 pc. The total number of cells
in the simulation is about $\simeq 4 \times 10^7$. About 25,000
timesteps have been performed for a total of 30,000 cpu hours.

To mimic a large scale converging 
flow (e.g. Audit \& Hennebelle 2005),
a converging velocity field is imposed at the left and right faces 
of the simulation box, on top of which velocity modulations have been 
superimposed. The velocity of each  incoming flow is  twice
the sound speed of the WNM, leading to a total velocity  difference of
about 40 km s$^{-1}$ within the box.
The amplitude of the modulation is about  a factor of two, and  it is
  periodic with a spatial frequency of 10 pc.
The boundary conditions are periodic for the 4 remaining faces.
Initially, the density and temperature are respectively 
1 cm$^{-3}$ and about 
8000 K, which  are also the values imposed at  the left and right faces.
The velocity is initially equal to zero through the box.
The magnetic field is uniform initially and parallel to the x-axis, therefore
aligned with the incoming velocity field and has an intensity of about
 5 $\mu$G, corresponding to equipartition between magnetic and thermal pressure
initially.
The cooling is due to atomic species as described in Audit \& Hennebelle 
(2005). Molecular cooling and H$_2$ formation are not treated at this stage.
As we will see, this nevertheless leads to  reasonable temperature and 
density  distributions. 

\setlength{\unitlength}{1cm}
\begin{figure*}
\begin{picture}(0,13.5)
\put(0,6.5){\includegraphics[width=7cm]{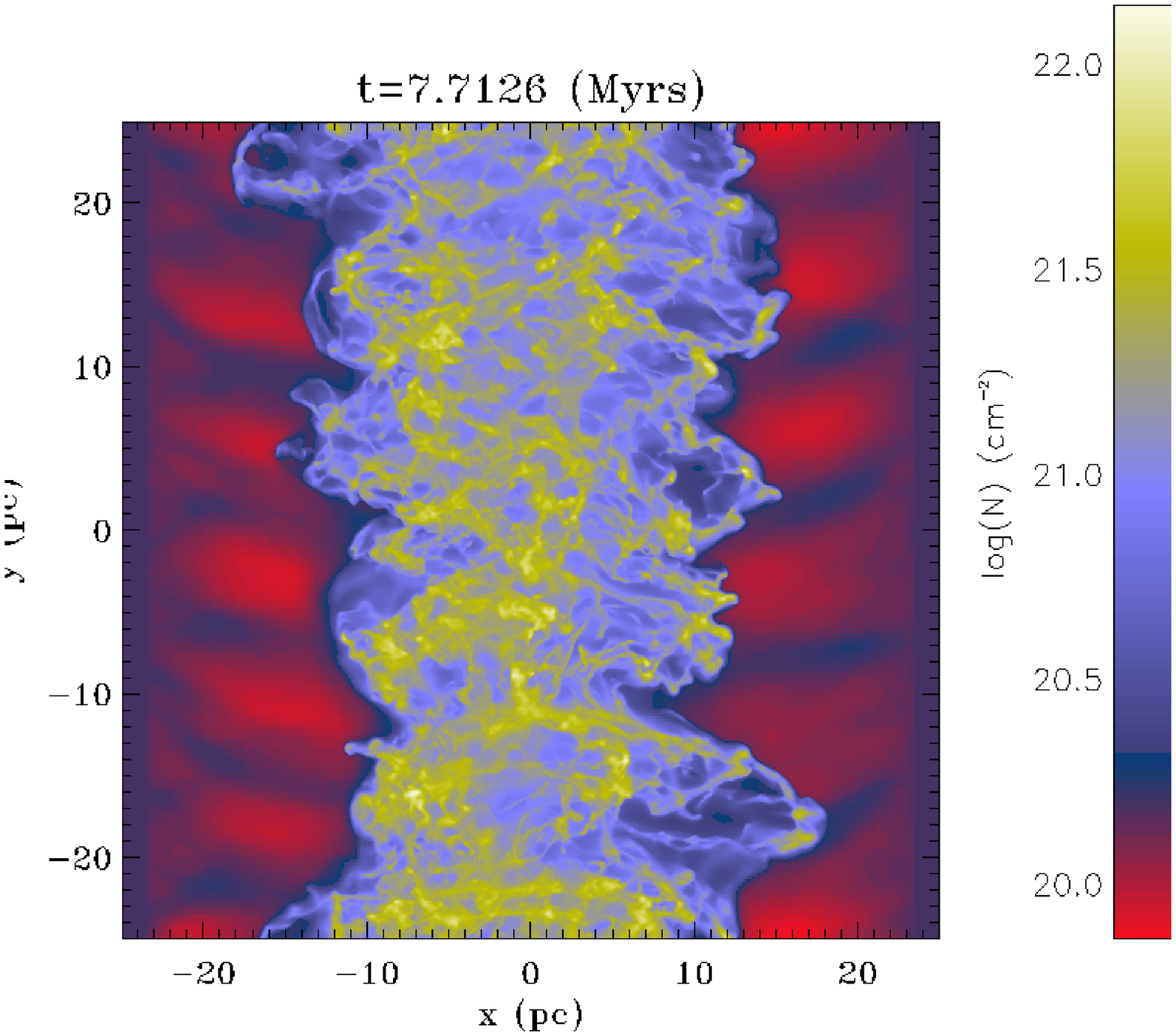}}
\put(0,0){\includegraphics[width=7cm]{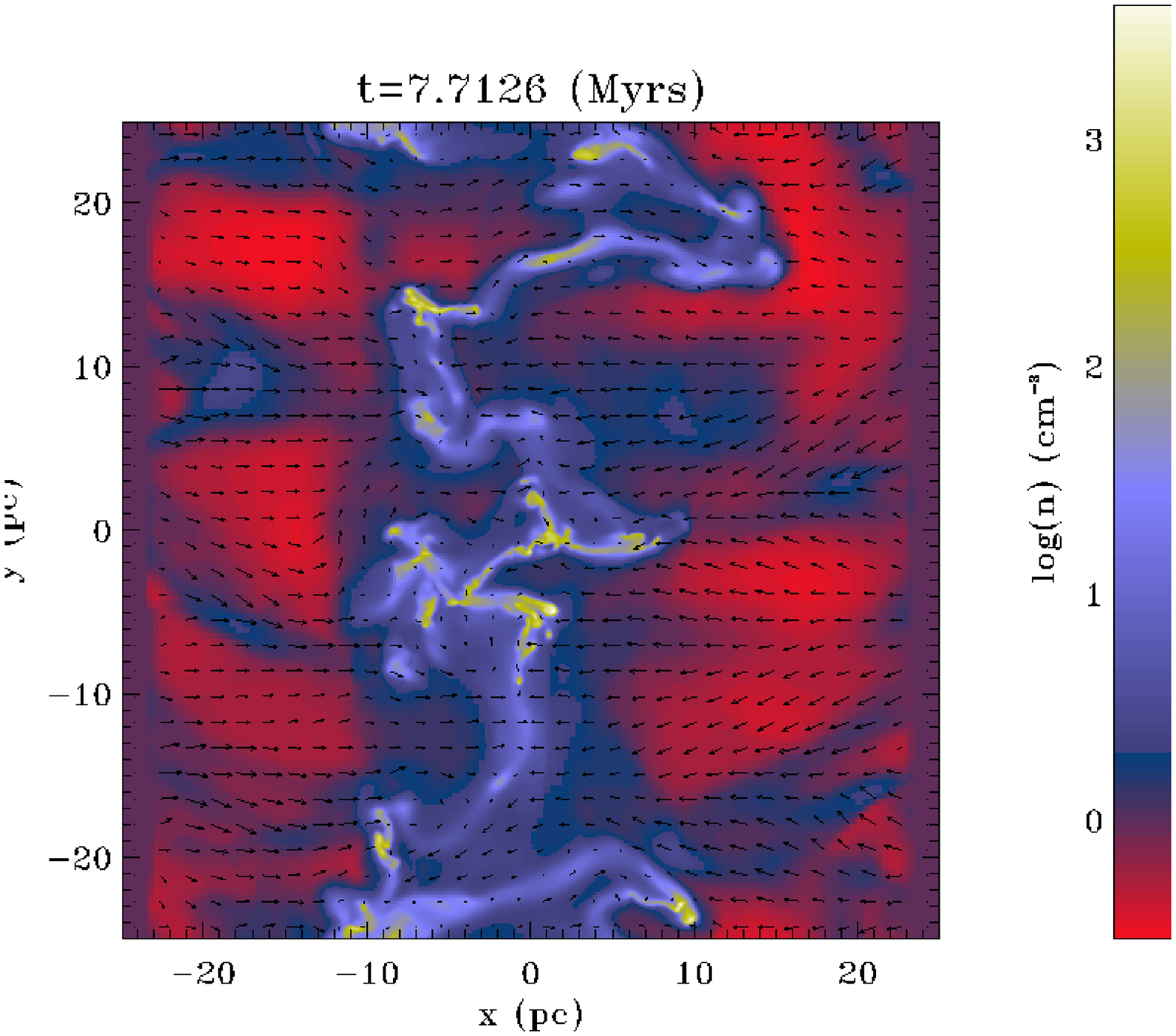}}
\put(9,6.5){\includegraphics[width=7cm]{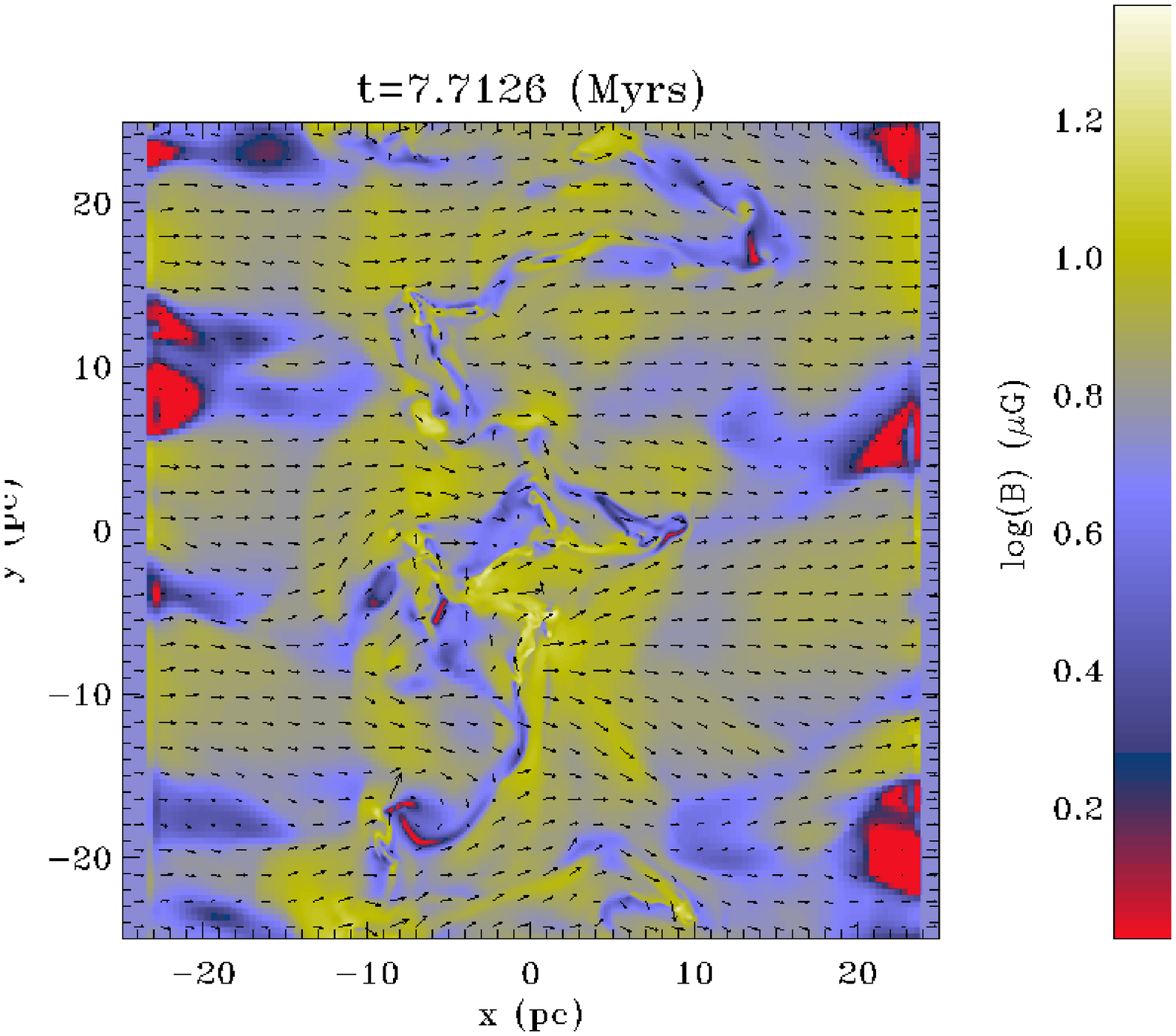}}
\put(9,0){\includegraphics[width=7cm]{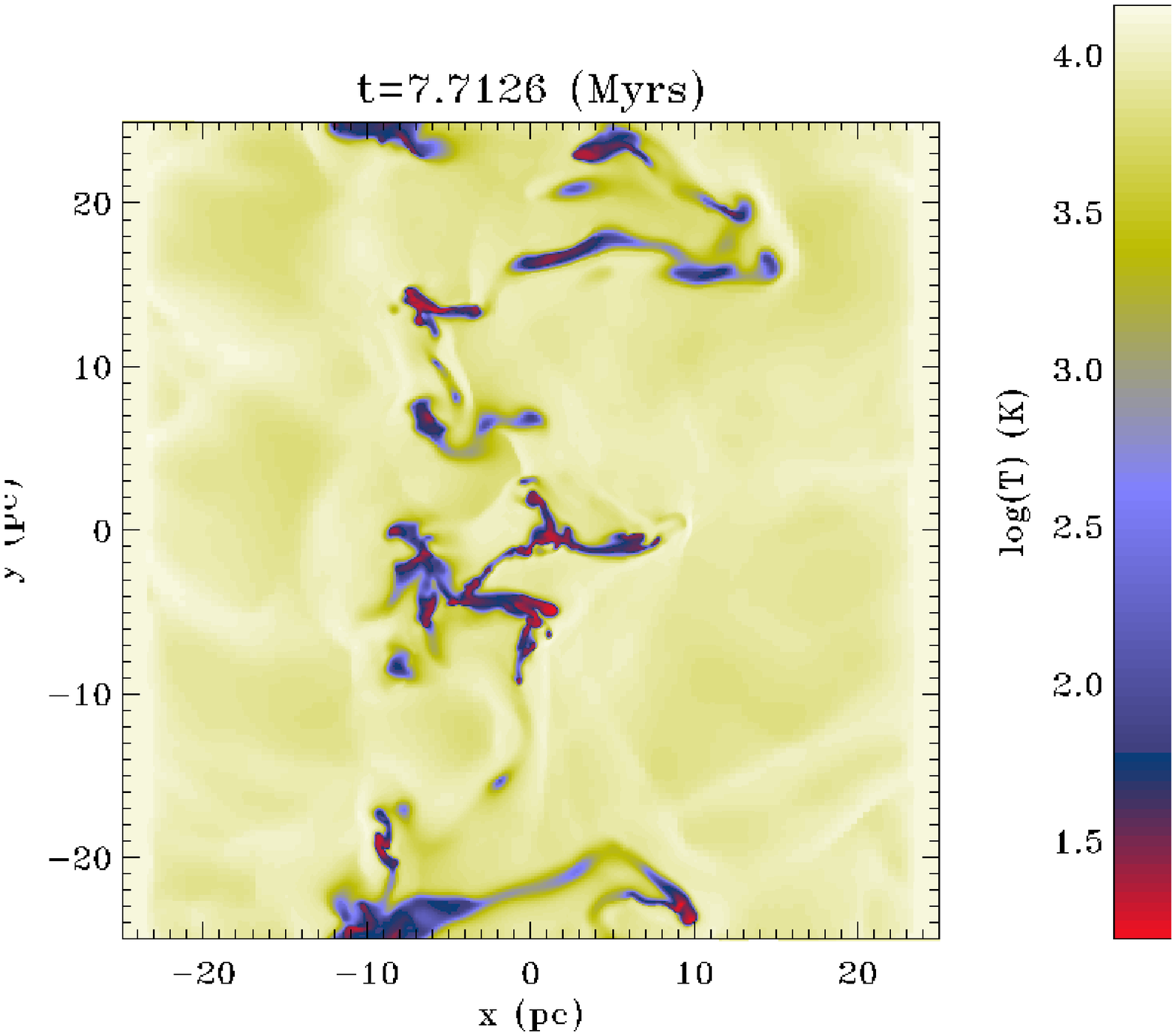}}
\end{picture}
\caption{Top left panel: column density. Top right panel:  Magnetic intensity 
and its xy-components (indicated as arrows) in the $z=0$ plane. Bottom left panel: density
  and velocity fields in the $z=0$ plane. Bottom right panel: temperature
 in the $z=0$ plane.}
\label{field}
\end{figure*}






\section{Results}
Figure~\ref{histo} shows the column density and density pdf as 
well as temperature, magnetic intensity and its variance as a function 
of density at time 3.71, 7.7, 12 and 13.79 Myr. Because of the mass 
injection, the total mass increases continuously within the simulation box 
from about 3000 $M_s$ initially to roughly 10 times this value at 
time 13.79 Myr. At time 3.71 and 7.7 Myr, the largest density reached in 
the simulation is  between a few times 10$^3$ cm$^{-3}$
 and $10^4 \pcc$, whereas at time 
13.79 Myr gravity has taken over and triggers gravitational 
collapse producing much higher densities. This indicates that the
cloud  should start
forming stars roughly 12 Myr after the collision of the converging flow 
has occurred ($t\simeq$ 1 Myr in the simulation).
At later times, the cloud keeps forming stars while the total mass continues
to increase. This is qualitatively in good agreement with the observations
reported by Blitz et al. (2007)  for the LMC, that the
masses of GMCs with little star formation activity are smaller than
those of GMCs with strong activity.
We note that the
duration of  7 Myr, they infer for the first  massive starless phase, is similar to the 
timescale of our simulation roughly estimated between time 7.7 Myr and 13.79
Myr. We stress, however, that precisely defining the
``birth'' time of the molecular cloud in our simulation is an elusive
task since the mass is an  increasing function of time. 
In future papers we will address
this point from an observational perspective. 

The  mass weighted PDF of the column density distribution peaks at about $2 \times 10^{21}$ cm$^{-2}$
and drops rapidly for higher values. It is interesting to note that
this is similar to what has recently been  inferred by Goldsmith et al. 
(2008) for the Taurus molecular cloud (see their figure 8).

The temperature drops  rapidly for densities between 
3 and 30 cm$^{-3}$ where it reaches a value of about 50 K. It then slowly 
decreases  down to about 10 K for densities higher than 10$^4$ cm$^{-3}$. 
Note that since UV shielding and molecular cooling are not considered here, 
 the temperature in the dense gas  is
probably overestimated. This would imply  
that the average density of the cold clumps should probably be a little
higher.  Explicitly treating
H$_2$ formation would have the same effect (Glover \& MacLow 2007).

For densities smaller than $\simeq 1000$ cm$^{-3}$, the magnetic 
intensity increases very smoothly  with density whereas 
for density larger than $\simeq 1000$ cm$^{-3}$, it is roughly 
proportional to $\sqrt{\rho}$.  Indeed,  the lower density gas is
  magnetically sub-critical and mainly flows along the magnetic field
  lines without compressing magnetic flux. On the other hand,  the 
 high density gas is supercritical and the magnetic flux is compressed along
with the dense gas under the influence of the gravitational force.
Future studies will have to investigate whether ambipolar diffusion modifies 
this behaviour significantly.
 The variance of the magnetic intensity increases 
smoothly  with density and ranges from about 
one third to half the mean magnetic intensity. We note that this 
magnetic intensity distribution appears to be very similar to what 
has been inferred from observations (Troland \& Heiles 1986, Crutcher 1999)

Figure~\ref{field} shows 
the  column density along the z-axis, the density and velocity field at 
$z=0$, and the temperature 
field in the same plane at time $t=7.7$ Myr after the beginning of 
the simulation. The column density  reveals that 
the cloud has a complex internal structure made of filaments and dense
clumps of density  between 100-1000 cm$^{-3}$,
 embedded  in a more diffuse phase. This is even more clearly 
apparent in the density  and temperature cuts which show 
that the clumps are relatively isolated and embedded into the warm diffuse
phase.
This suggests that the molecular clouds are not
homogeneous isothermal media,
but rather, consist of a complex array of
filaments and clumps, with denser gas being colder, and therefore
with little or no excess of thermal pressure over the surrounding gas. 

Interestingly, the density of the 
warm gas embedded in the molecular cloud
 is higher than  in the outer medium ($n \simeq 1 $ cm$^{-3}$)
 and can be as large as 3-4 cm$^{-3}$. 
Indeed,  this gas has been previously shocked 
 and it is in the process of cooling as it moves towards
the dense cold regions (\VS\ et al. 2006; Hennebelle \& Audit 2007).
From a comparison between the column density and temperature distribution 
in the $z=0$ plane, 
we note that the warm gas is deeply embedded in the molecular cloud.  
Note that in this work, the UV field is assumed to be constant.
 Although this is obviously not a good assumption for the 
dense gas, we see that since the filling factor of the cloud appears to be 
small, this is certainly a fair assumption  for the WNM even when it is 
deeply embedded inside the cloud. Moreover, 
 the higher temperatures are sometimes found at the edge of the clumps 
at the onset of the accretion shocks which occurs when the WNM flow 
encounters a dense clump. This clearly indicates that 
the dissipation of mechanical energy plays an active role in the heating 
of the warm phase. Altogether, this is in good agreement with the 
picture proposed by Hennebelle \& Inutsuka (2006) except that the 
mechanical energy which heats the warm phase is the kinetic energy of
the shocks rather than the energy of the MHD waves (possibly underestimated 
in this work since the ion-neutral drift is not treated). 
Note that the finding of interclump medium being low density atomic hydrogen
($n<4-10$ cm$^{-3}$) is consistent with the estimate of Williams et al. (1995) 
 for the Rosette molecular cloud.

Figure~\ref{field} also shows the magnetic intensity in the $z=0$ plane
 as well 
as its xy-components. In the external medium the magnetic field remains
much more uniform than in the dense regions, where its direction 
fluctuates significantly. 
Since the field was initially
uniform, this implies that the turbulent
motions are able to significantly distort the field.

\section{Conclusion}

We have presented the results of AMR MHD simulations aiming to
describe self-consistently the formation of a molecular cloud from a
converging flow of warm diffuse atomic hydrogen.  Our simulations
suggest a complex multi-phase structure of the molecular clouds which
consists  of cold and dense clumps embedded into a warm atomic phase.
The simulations reproduce reasonably well the observed variations of magnetic
  intensity with density and column density distribution.
Finally, we suggest that star
formation may start in the cloud while it is still accreting
material. This would imply that the mass of GMCs vary with
time, in
good agreement with recent observations (Blitz et al. 2007), and with the
results of previous non-magnetic simulations of the phenomenon.

\end{document}